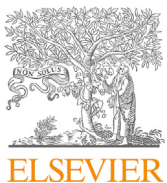

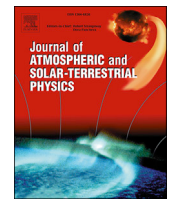

# Nighttime mesospheric ozone enhancements during the 2002 southern hemispheric major stratospheric warming

Christine Smith-Johnsen [a, *], Yvan Orsolini [b,c], Frode Stordal [a], Varavut Limpasuvan [d], Kristell Pérot [e]

[a] Department of Geosciences, University of Oslo, Norway
[b] Birkeland Centre for Space Science, University of Bergen, Norway
[c] Norwegian Institute for Air Research, Kjeller, Norway
[d] Department of Coastal and Marine Systems Science, Coastal Carolina University, Conway, SC, USA
[e] Department of Earth and Space Sciences, Chalmers University of Technology, Gothenburg, Sweden

ABSTRACT

Sudden Stratospheric Warmings (SSW) affect the chemistry and dynamics of the middle atmosphere. Major warmings occur roughly every second winter in the Northern Hemisphere (NH), but has only been observed once in the Southern Hemisphere (SH), during the Antarctic winter of 2002. Observations by the Global Ozone Monitoring by Occultation of Stars (GOMOS, an instrument on board Envisat) during this rare event, show a 40% increase of ozone in the nighttime secondary ozone layer at subpolar latitudes compared to non-SSW years. This study investigates the cause of the mesospheric nighttime ozone increase, using the National Center for Atmospheric Research (NCAR) Whole Atmosphere Community Climate Model with specified dynamics (SD-WACCM). The 2002 SH winter was characterized by several reductions of the strength of the polar night jet in the upper stratosphere before the jet reversed completely, marking the onset of the major SSW. At the time of these wind reductions, corresponding episodic increases can be seen in the modelled nighttime secondary ozone layer. This ozone increase is attributed largely to enhanced upwelling and the associated cooling of the altitude region in conjunction with the wind reversal. This is in correspondence to similar studies of SSW induced ozone enhancements in NH. But unlike its NH counterpart, the SH secondary ozone layer appeared to be impacted less by episodic variations in atomic hydrogen. Seasonally decreasing atomic hydrogen plays however a larger role in SH compared to NH.

## 1. Introduction

Major Sudden Stratospheric Warmings (SSWs) are common in the Northern Hemisphere (NH), where they occur about every second winter (Charlton and Polvani, 2007), though with large inter-annual and decadal variability. In the Southern Hemisphere (SH), only one major SSW has ever been observed since the discovery of SSWs in the 1950s (Scherhag, 1952), and that was during the Antarctic winter of 2002. According to the World Meteorological Organization (WMO) definition (McInturf, 1978), a SSW is major when the Polar Night Jet (PNJ) reverses from eastward to westward at 60° latitude at the 10 hPa pressure level, and the meridional temperature gradient between 60 and 90° latitude also reverses. Reductions of the zonal-mean zonal wind occur frequently in the upper stratosphere and mesosphere without propagating down to 10 hPa.

The changes in the large scale circulation are triggered by the amplified stratospheric planetary waves during a major SSW event, which also alter the vertical propagation of gravity waves. The westward PNJ during a major SSW prevents gravity waves with westward phase speeds from propagating up from the troposphere into the mesosphere. During normal winter conditions, such gravity waves dissipate in the mesosphere and lower thermosphere, exerting a westward drag on the background wind and downwelling in polar regions. With the PNJ reversal throughout the stratosphere, only gravity waves with eastward phase speeds (exceeding the tropospheric eastward wind speed) are allowed to propagate up into the mesosphere. The eastward gravity waves reduce the normal westward wave drag, and thus contribute to an anomalous ascent and cooling in the mesosphere (Siskind et al., 2005;

Limpasuvan et al., 2012, 2016; Funke et al., 2017). In addition to the warming of the polar stratosphere and the cooling of the polar mesosphere, there are also model and observational evidences of a lower thermospheric warming following major SSWs (Funke et al., 2010; Liu and Roble, 2002).

In the mesopause region (at 85–95 km altitude), high above the main stratospheric ozone layer at 30 km altitude, a secondary peak in the ozone density is found (Hays and Roble, 1973; Kaufmann et al., 2003), due to the cold temperature and the high atomic oxygen density (Smith and Marsh, 2005). The main nighttime ozone source at this altitude is atomic oxygen, while its sinks are atomic hydrogen and atomic oxygen. During daytime, photolysis is the major loss process. The nighttime ozone in this secondary layer is in photochemical equilibrium (Smith and Marsh, 2005), and the ozone density (given as molecules/cm$^3$) can be expressed as:

$$[O_3] = \frac{c_1 \cdot [O] \cdot [O_2] \cdot M}{c_2 \cdot [H] + c_3 \cdot [O]} \tag{1}$$

The reactions rates ($c_1$ to $c_3$) and the air density M are all temperature dependent, with $c_1 = 6 \cdot 10^{-34} (300/T)^{2.4}$ $cm^6 molecules^{-2} s^{-1}$, $c_2 = 1.4 \cdot 10^{-10} exp(-470/T)$, $c_3 = 8 \cdot 10^{-12} exp(-2060/T)$ $cm^3 molecules^{-1} s^{-1}$, and $M = p/kT$ (where $p$ is pressure and $k$ is Boltzmann's constant) (Sander et al., 2006). From the reaction rates, it can be seen that the ozone abundance is inversely related to temperature. Because ozone has a very short lifetime (on the order of minutes), ozone produced at night is not subject to transport (Smith and Marsh, 2005). Rather, the local concentration rapidly responds to transport of the longer lived sources and sinks of ozone. At daytime, ozone is quickly destroyed by sunlight, and the concentrations are orders of magnitude lower than at night.

(Tweedy et al., 2013) showed through a combined model and observational study that a brief enhancement of the nighttime ozone occurs in the secondary layer at high latitudes during the NH major SSWs. The ozone enhancement is mostly driven by two factors, primarily by the mesospheric cooling, and to a lesser extent by the decrease in atomic hydrogen. There is also a small negative contribution from atomic oxygen to counteract the ozone increase.

During the Antarctic winter of 2002, several precursory mesospheric dynamical perturbations occurred prior to the major SSW, documented by ground-based radar wind or airglow measurements (Espy et al., 2005; Azeem et al., 2010). Model (Coy et al., 2005) studies also show mesospheric coolings associated with the zonal-mean zonal wind decreases.

The aim of this study is to further investigate the impact on the secondary ozone layer of the SH major warming of 2002 and its precursory warmings. While this event has been extensively described in the literature (e.g (Shepherd et al., 2005)) and in particular its influence upon the Antarctic ozone hole, there has been no study to date on its impact on the secondary ozone layer. Our results will be compared to (Tweedy et al., 2013) to determine if there are inter-hemispherical differences in the secondary ozone layer response to SSWs.

## 2. Methods, model and satellite observations

As a motivation for this study of the secondary ozone layer during the Antarctic winter and spring 2002, we have used ozone observations by the Global Ozone Monitoring by Occultation of Stars (GOMOS). GOMOS is a stellar occultation instrument on board the European platform ENVISAT, launched in March 2002 and in operation until April 2012 (see (Bertaux et al., 2010) for an overview of the mission). Ozone profiles are retrieved from measurements in the UV-visible spectral range, using the stellar occultation method, from the troposphere to the lower thermosphere. The vertical resolution in the mesosphere is 3 km, and the observations reach up to about 105 km altitude. The retrieval technique is described by (Kyrölä et al., 2010) and an analysis of retrieval errors is presented in (Tamminen et al., 2010). Only nighttime measurements are used. To ensure a good quality of the mesospheric ozone profiles, the data has been filtered based on the star effective temperature. Following European Space Agency's recommendation (European Space Agency, 2007), we screen out all profiles from cold stars with temperatures lower than 6000 K.

GOMOS regular observations began in mid-August 2002, so the September ozone profiles in our study is based on the very first measurements of the mission, which have been hardly used before. In September 2002, GOMOS was measuring mostly before sunrise (between 3AM and 6 AM local time), in the latitude range 55–65°S. The number of observations during this starting period is also limited, and due to these restrictions, only a September mean of the GOMOS ozone profiles is used. For comparison with the September 2002 monthly-mean ozone profile, we use the September mean from the two following years. GOMOS was affected by a technical problem in 2005, which resulted in a change in the occultation scheme after that year. Hence, to ensure a temporal and geographical sampling similar to 2002, we use only the measurements made before this malfunction. The ozone measurements from the years 2003 and 2004 could be influenced by the high geomagnetic activity level (Andersson et al., 2014), which would result in lower ozone level in these two years. The climatology we use could thus be an underestimation.

To further examine the secondary ozone layer during the major SSW, the National Center for Atmospheric Research (NCAR) Whole Atmosphere Community Climate Model (WACCM 4) is used. WACCM is a chemistry-climate model that reaches up to 140 km altitude, and is part of the Community Earth System Model (CESM) (Marsh et al., 2013). WACCM is used with specified dynamics (SD-WACCM), where the horizontal winds, temperature and surface pressure are constrained to analyses from NASA Global Modeling and Assimilation Office Modern-Era Retrospective Analysis for Research and Applications (MERRA) (Rienecker et al., 2011), by the method described in (Kunz et al., 2011). This nudging is applied from the surface to about 50 km altitude (0.79 hPa), and the model is free running above 60 km (0.19 hPa), with a linear transition in between. The model has a horizontal resolution of 1.9° latitude and 2.5° longitude, and a variable vertical resolution with 88 vertical levels. The model has been run from 1990 until 2010, with output once a day (00:00 GMT). This sampling will cause our nighttime averages to always cover the Atlantic hemisphere, and not the Indian/-Pacific. We examine in detail the Antarctic winter of 2002, and compare it to the model climatology over the years 1990–2010 (excluding 2002). The diagnostics include the Transformed Eulerian Mean (TEM) meridional circulation.

It is worth noting that WACCM tends to underestimate the ozone nighttime abundance in the secondary layer. Comparing ozone from WACCM with measurements by the Sounding of the Atmosphere using Broadband Emission Radiometry (SABER) on the Thermosphere-Ionosphere-Mesosphere Energetics and Dynamics (TIMED) satellite (Smith et al., 2015), showed that ozone volume mixing ratios were 50 percent lower than observed. They pointed out that this difference is likely due to a negative bias in atomic oxygen, as too little atomic oxygen is transported down from the lower thermosphere into the mesosphere as a result of weak vertical diffusion and eddy mixing in the model. Since we are interested in ozone *variability* in this paper, the model bias is not so critical.

## 3. Results and discussion

Applying the WMO definition of a SSW to SD-WACCM, which is nearly identical to MERRA analyses at 10 hPa, the reversal defining the onset date occurred on 25 September 2002. The minimum in the zonal-mean wind (hence when most westward) was found two days after, on September 27. As mentioned in the Introduction, temporary reductions in the zonal-mean zonal wind occur frequently in the upper stratosphere and mesosphere, but do not always propagate down to 10 hPa. Since our focus is on the mesosphere, we use the SD-WACCM zonal-mean zonal wind at the 1 hPa pressure level to identify not only the major warming but also the minor ones at higher altitudes. Using zonal wind reversal at 1 hPa to identify major SSWs was also done by (Limpasuvan et al., 2016). Fig. 1 shows three reductions of the PNJ earlier in the 2002 SH winter, before the strongest reduction that leads to a full zonal wind reversal to westward, that will ultimately propagate down to 10 hPa and define the SSW onset. In the following figures, these three wind reductions and the full reversal will be highlighted by vertical lines. Similar oscillations are observed earlier in the Antarctic summer (not shown). However, the three episodes shown in Fig. 1 are associated with the strongest and clearest signals in both ozone and temperature. For presentation, we limit the time period shown to be between August 15 and October 15.

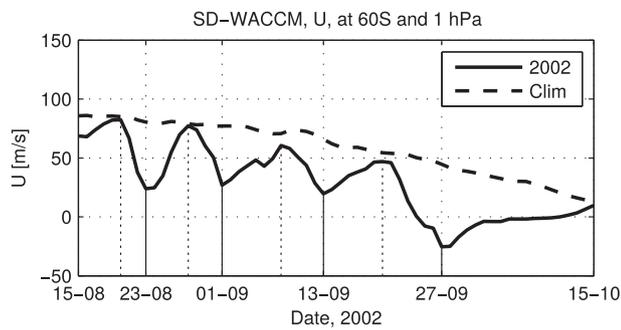

**Fig. 1.** Zonal-mean of zonal wind U (m/s) from SD-WACCM at 1 hPa pressure level and 60°S latitude in 2002, contrasted by the 1990–2010 climatology (dashed line). Three reductions in the polar night jet are seen, before it reverses on September 25, 2002. The full vertical lines highlights the timing of the local minimum wind speed, and the dashed lines the timing of the local maximum wind speed.

The September mean ozone from GOMOS in 2002 shows a 40% ozone increase at mesospheric altitudes compared to the corresponding mean of the next two years (Fig. 2). A similar relative increase is seen in the secondary ozone layer modeled by SD-WACCM (Fig. 2), although the model significantly underestimates ozone due to the previously mentioned negative bias. The GOMOS measurements are sparse on any given day in the period considered, fluctuating between between 2 and 40 measurements per day in September. There is also a 1-week gap after

September 8. For those reasons, we have not carried out a spatial collocation when comparing GOMOS directly to SD-WACCM. But we do see an increase both in the observed and modeled nighttime mesospheric ozone, and we will carry out the rest of our study using the model.

The zonal-mean zonal winds at 60°S in 2002 can be compared to the climatology over the period 1990–2010 (Fig. 3, top row). The climatological eastward PNJ stretches from 20 km to 80 km during SH winters, while westward winds prevail above. The eastward winds progressively weaken during September and early October, while westward winds descend to the mesosphere. In the winter of 2002, the PNJ weakened and then recovered several times near its peak altitude near the stratopause (1 hPa), as seen by the dashed vertical lines in Fig. 1. These three reductions at 1 hPa actually correspond to full reversals to westward at higher altitudes. In late September however, the wind is reversed to westward and the reversal propagated down to 1 hPa, and then further downwards into the stratosphere. After the late September reversal, the PNJ recovered to eastward only below 30 km and above 60 km, while the westward regime persists in the altitudes between 30 and 60 km.

Concomitant to the PNJ reversals, changes occur in the vertical component of the TEM circulation (w*) in the polar vortex (see Fig. 3, second row). A strong mesosphere descent prevails until mid-September in the climatology, followed by a weak ascent during the spring transition in late September and early October. The anomalous ascent (i.e., departure from climatology) is seen in the mesosphere associated with each of the four PNJ reductions, albeit at slightly different altitudes: near 80–90 km during the first PNJ reduction, and at 70–80 km for the three last reductions. Small fluctuations can also be seen when no PNJ reductions are present. The upwelling seen at the time of the SSW onset covers the largest vertical range (40–80 km), and lasts into October. The strong descent in the upper mesosphere above 80 km is enhanced after the onset, as was observed in WACCM simulations during NH SSWs by (Kvissel et al., 2012; Limpasuvan et al., 2016). Also before the SSW onset, an anomalously descent is found above 80 km, contributing to warmer temperatures around the mesopause.

The zonal-mean temperature within the polar vortex shows warming in the stratosphere at the time of the downwelling (third row of Fig. 3). Repeated anomalous cooling episodes can be seen below the cold mesopause region at the time of the wind reductions and ascent, showing a downward propagation with time. The SSW onset gives rise to the

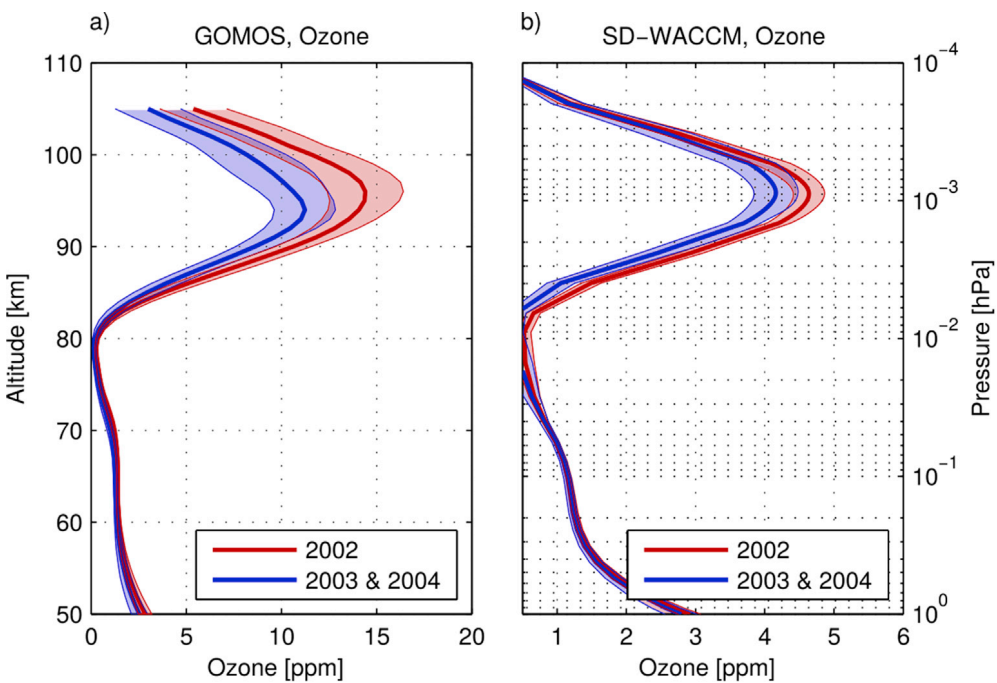

**Fig. 2.** a) GOMOS ozone [ppm] profile, September monthly mean from 2002 (red line), contrasted by the September monthly mean from 2003 to 2004 (blue line). The 2002 mean is an average of 439 profiles at 55-65°S latitude. b) WACCM ozone [ppm] profile, September monthly mean over latitudes 55-65°S from 2002 (red line), contrasted by the September monthly mean from 2003 to 2004 (blue line). Not the x-axis is not the same in the two panels, as WACCM is known to underestimate ozone at these altitudes (Smith et al., 2015). (For interpretation of the references to colour in this figure legend, the reader is referred to the Web version of this article.)

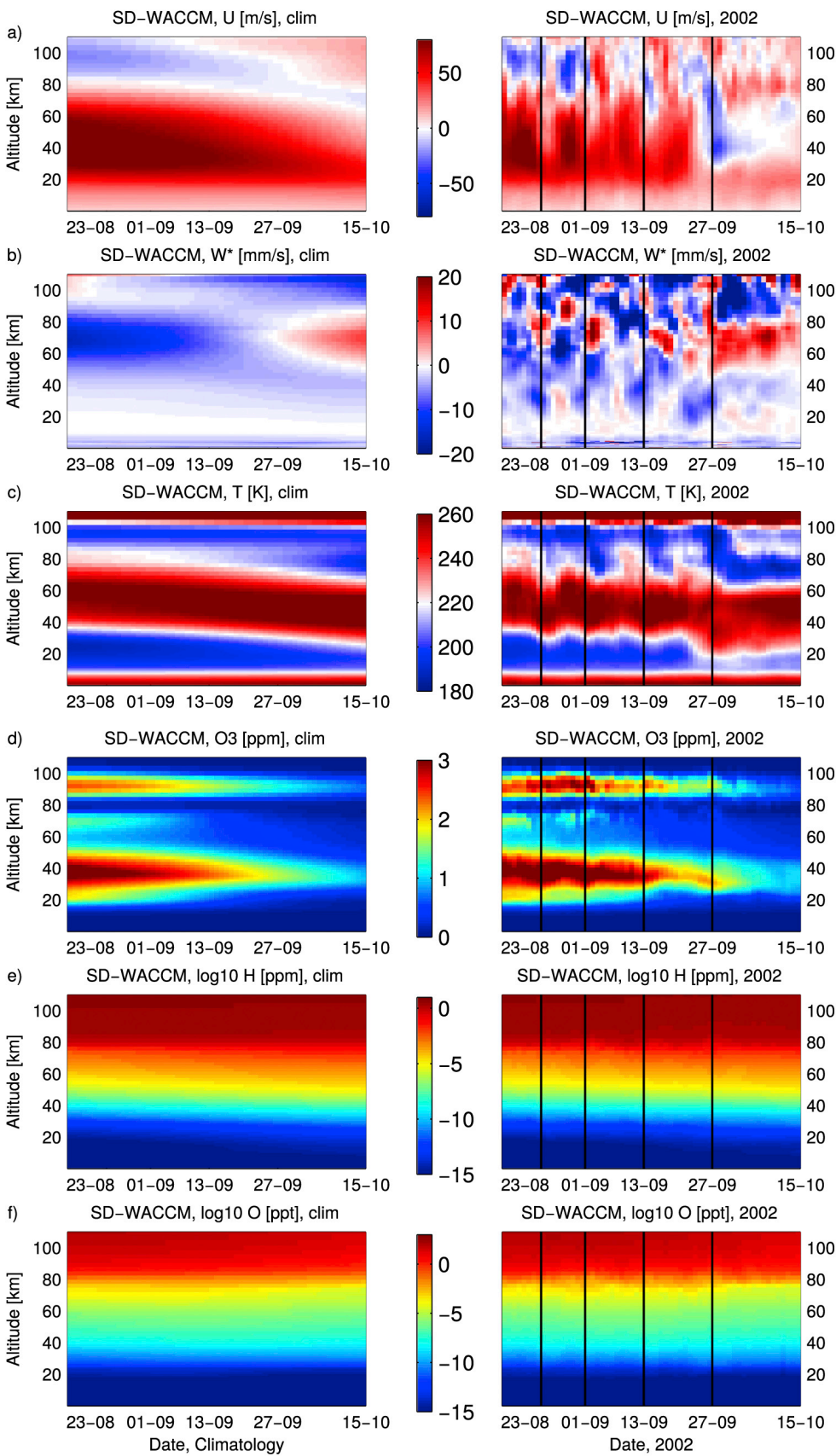

**Fig. 3.** Zonal mean of: a) zonal wind U (m/s), b) vertical component of the residual mean circulation w* (mm/s), c) temperature (K), d) ozone (ppm), e) atomic hydrogen (log10 ppm), f) atomic oxygen (log10 ppt) from SD-WACCM. The panels to the left are the climatologies from 1990 to 2010. Panels to the right are from 2002, where the reductions in polar night jet are highlighted by vertical lines. The zonal wind is a latitude mean around 60° latitude to capture the polar night jet, while the other parameters are averaged over latitudes 60°S - 90°S to cover the southern hemispheric polar vortex.

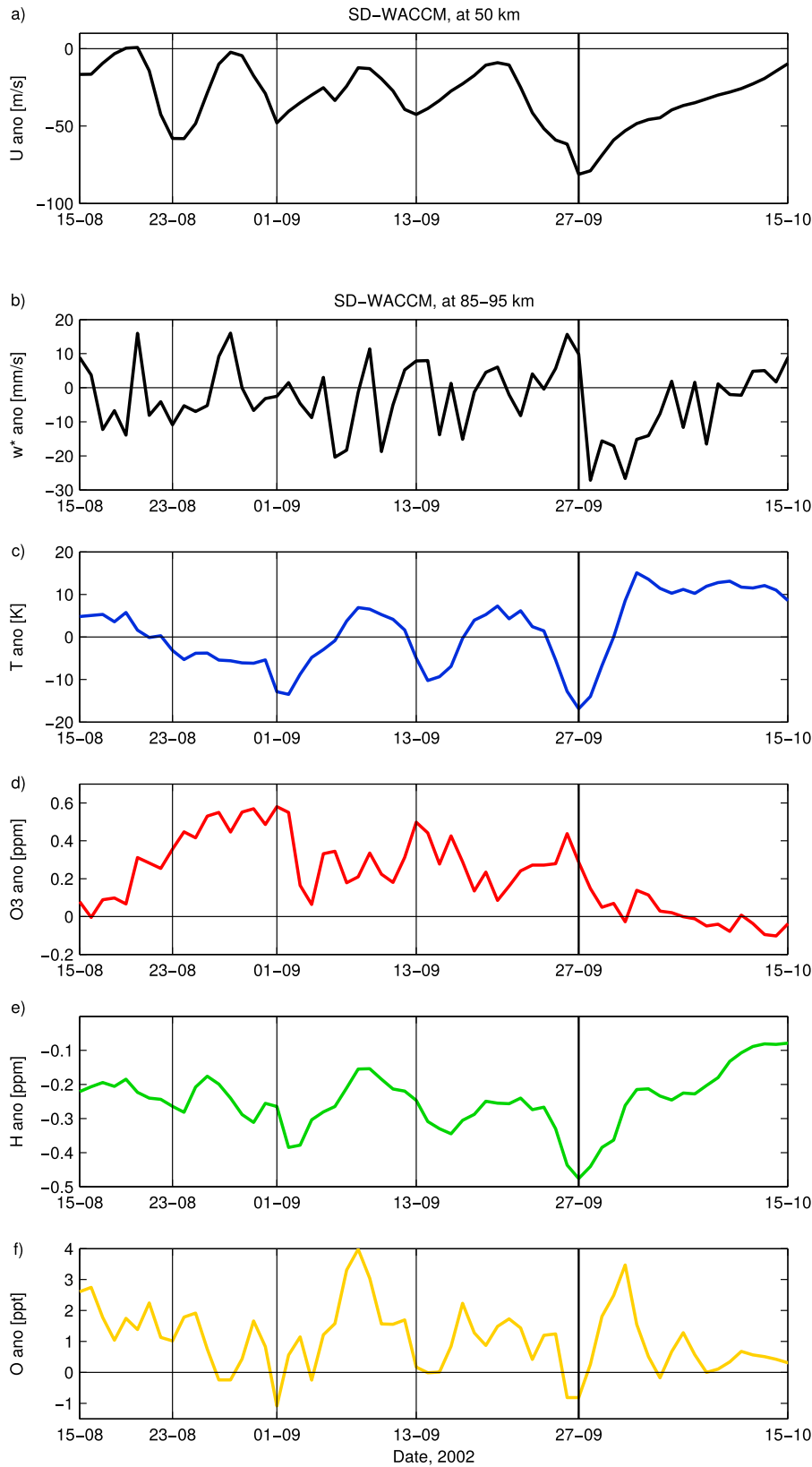

**Fig. 4.** Nighttime polar cap mean over latitudes 60°S - 90°S, of: a) zonal wind U (m/s), b) vertical component of the residual mean circulation w* (mm/s), c) temperature (K), d) ozone (ppm), e) atomic hydrogen (log10 ppm), f) atomic oxygen (log10 ppt), All panels show anomalies from the 1990–2010 climatology. The altitude shown is 85–95 km, where the secondary ozone layer is situated, for all parameters except the zonal wind which is at 50 km altitude.

strongest and most persistent cooling, starting at 100 km and propagating down to 60 km. The strong upper mesospheric descent then starts at the same time above 80 km, causing a warm anomaly. We note that during the minor warming on 23rd of August 2002, the period of mesospheric cooling was prolonged, lasting over a week. Using the NOGAPS (Navy Operational Global Atmospheric Prediction System) model and observations from SABER (Coy et al., 2005), found a stratospheric warming and a mesospheric cooling at the onset of the minor warming event,

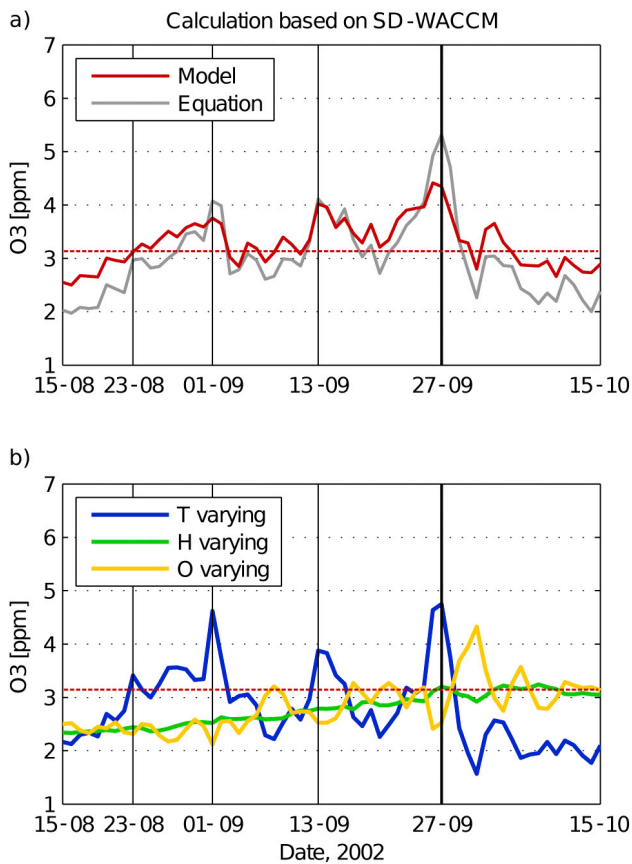

a)

**Calculation based on SD-WACCM**

b)

**Fig. 5.** a): Nighttime polar cap mean over latitudes 60°-90°S, of ozone (ppm) at 85–95 km altitude from SD-WACCM (red). The gray line shows nighttime ozone calculated from the equation for chemical equilibrium, Equation (1), based on temperature, atomic hydrogen and atomic oxygen from SD-WACCM. The horizontal dashed red line is the mean ozone from WACCM for this time period. b): Ozone (ppm) calculated from the chemical equilibrium equation, with only one of the constituents varying at a time. The blue line shows the ozone caused by variations in temperature, while hydrogen and oxygen are fixed to their mean value for the time period. The green line shows ozone based on atomic hydrogen variations, and the yellow line shows ozone for varying atomic oxygen. The horizontal dashed red line is the mean ozone from WACCM for this time period. (For interpretation of the references to colour in this figure legend, the reader is referred to the Web version of this article.)

associated to stratospheric downwelling and mesospheric upwelling, respectively.

The secondary ozone layer is situated in the mesopause region around 85–95 km (fourth row in Fig. 3). As mentioned in the Introduction, the ozone has a much higher volume mixing ratio at night time compared to daytime. This explains why the polar-averaged climatological ozone in Fig. 3 decreases towards spring, since the night shortens and insolation increases. Corresponding to the four wind reductions and mesospheric coolings in August and September 2002, increases can be seen in the secondary ozone layer. By early October, the upper mesosphere warming associated with the strong descent contributes to lower ozone in the secondary layer, as expected from the ozone and temperature anti-correlation. To show these fluctuations more clearly, we now examine ozone anomalies from climatology and use nighttime means only.

Both atomic hydrogen and atomic oxygen (fifth row and sixth row in Fig. 3) increase exponentially with increasing altitude. At the altitude of the secondary ozone layer, they both show a seasonal decrease from August towards October. At the time of the wind reductions in 2002, episodic decreases are seen in the atomic oxygen and hydrogen (this is seen clearer from the anomalies in Fig. 4).

Fig. 4 shows the nighttime polar ozone averaged over 85–95 km along with corresponding time series of zonal wind, vertical wind, temperature, atomic hydrogen and atomic oxygen. The figure reveals how well the positive ozone anomalies correspond to temperature negative anomalies at the time of the wind reversals, reflecting a negative correlation between temperature and ozone (correlation coefficient of −0.8). The event starting on August 23 shows a period of prolonged cooling and ozone increase, that culminates in a short-lived temperature minimum and ozone peak in the first days of September. This event is hence of different characteristic from the others, and the August 23 and September 1 event are not necessarily completely separate events.

A negative anomaly in hydrogen indicates that the amount of atomic hydrogen (the main ozone sink) is lower in 2002 than in the climatology throughout the period. In addition, atomic hydrogen shows decreases that follow the PNJ reductions. Atomic oxygen, which is both a source and a sink for ozone, also show negative anomalies around the time of the ozone increases. The largest decrease in temperature, atomic hydrogen and atomic oxygen occurs on the SSW onset, corresponding to the strong upwelling. As the strong upper mesospheric descent later takes over in the days following the SSW, there is a strong increase in temperature, as mentioned earlier, but also in atomic hydrogen and atomic oxygen.

Nighttime ozone in the mesosphere is in chemical equilibrium and only depends on temperature, on the concentrations of atomic and molecular oxygen, and of atomic hydrogen, as given by Equation (1). In order to demonstrate that the modeled and the equilibrium ozone are in agreement throughout the period, we use the local temperature and densities of atomic hydrogen and atomic oxygen from SD-WACCM to calculate the nighttime ozone chemical equilibrium in Fig. 5. This equilibrium relation can be used to understand the relative importance of the factors for the ozone photochemical equilibrium (i.e. temperature, atomic hydrogen, or atomic oxygen). Following the same procedure as (Tweedy et al., 2013), we let one of these factors vary at a time, while the other factors are fixed to their mean values over the studied period. For example, to see the effect of temperature upon nighttime ozone, Equation (1) is used with time-mean values for atomic hydrogen and atomic oxygen, while only the nighttime temperature is varied. This is shown as the blue line in Fig. 5. The green and yellow coloured curves indicate the contributions of atomic hydrogen and oxygen respectively, again keeping the other two factors constant. Similar to what was found by (Tweedy et al., 2013) for the NH SSWs, the temperature is found to be the main contributor to ozone changes. The temperature-only contribution gives even higher ozone increases than what is found in the model, but this is counteracted by atomic oxygen at these times. During the upper mesospheric descent that follows the SSW onset, ozone is declining due to the warmer temperatures. This effect is counteracted by the brief pulse of atomic oxygen early in October. The atomic hydrogen decreases during the wind reductions only contribute weakly to the ozone increase. However, the slow seasonal decrease of atomic hydrogen (as seen in Fig. 3), would alone cause a slow build-up of the ozone from mid-winter toward summer.

In summary, there is a good correspondence with the NH results in (Tweedy et al., 2013), who found that temperature was the driving factor behind ozone increases. During the NH mesospheric coolings associated to SSWs studied by (Tweedy et al., 2013), the mean temperature decrease was about 20 K, while it is somewhat weaker here (about 15 K), and the associated mean response in ozone was stronger (an increase of 2.5 ppm, while we see 0.7 ppm). There is however a difference regarding atomic hydrogen, which (Tweedy et al., 2013) found was the 2nd main factor in the NH. Here, by contrast, atomic hydrogen has little impact on ozone during the minor and major warming events (as idicated by the lack of fluctuations in the green line in Fig. 5). However, the influence by the seasonal cycle of atomic hydrogen in the SH is more pronounced (see the slow build-up in the green line in Fig. 5) than what is found in the NH (Tweedy et al., 2013) (Smith et al., 2015) studied the climatology, seasonal cycle and the year-to-year variability of upper mesospheric ozone, temperature, atomic hydrogen and atomic oxygen, at high latitudes of both hemispheres, comparing SD-WACCM to SABER satellite observations. Incidentally, they also see the ozone increase at the time of the SH SSW (see their Fig. 9). That study linked the seasonal decline of atomic hydrogen in late austral winter to the weakening of the mean meridional circulation, which then ceases to bring hydrogen-rich air from lower latitudes. The SH events investigated here occurred in late winter and early spring when atomic hydrogen is declining, while the NH mid-winter events investigated by (Tweedy et al., 2013) occurred when it is increasing. The amplitude of the atomic hydrogen fluctuations in (Tweedy et al., 2013) were 3 times larger (their Fig. 5). The occurrence in 2002 at a period of low solar activity might also explain why the background hydrogen is lower consistently throughout the winter 2002. In our case, there is also a small negative (counter-acting the ozone increase) contribution from atomic oxygen.

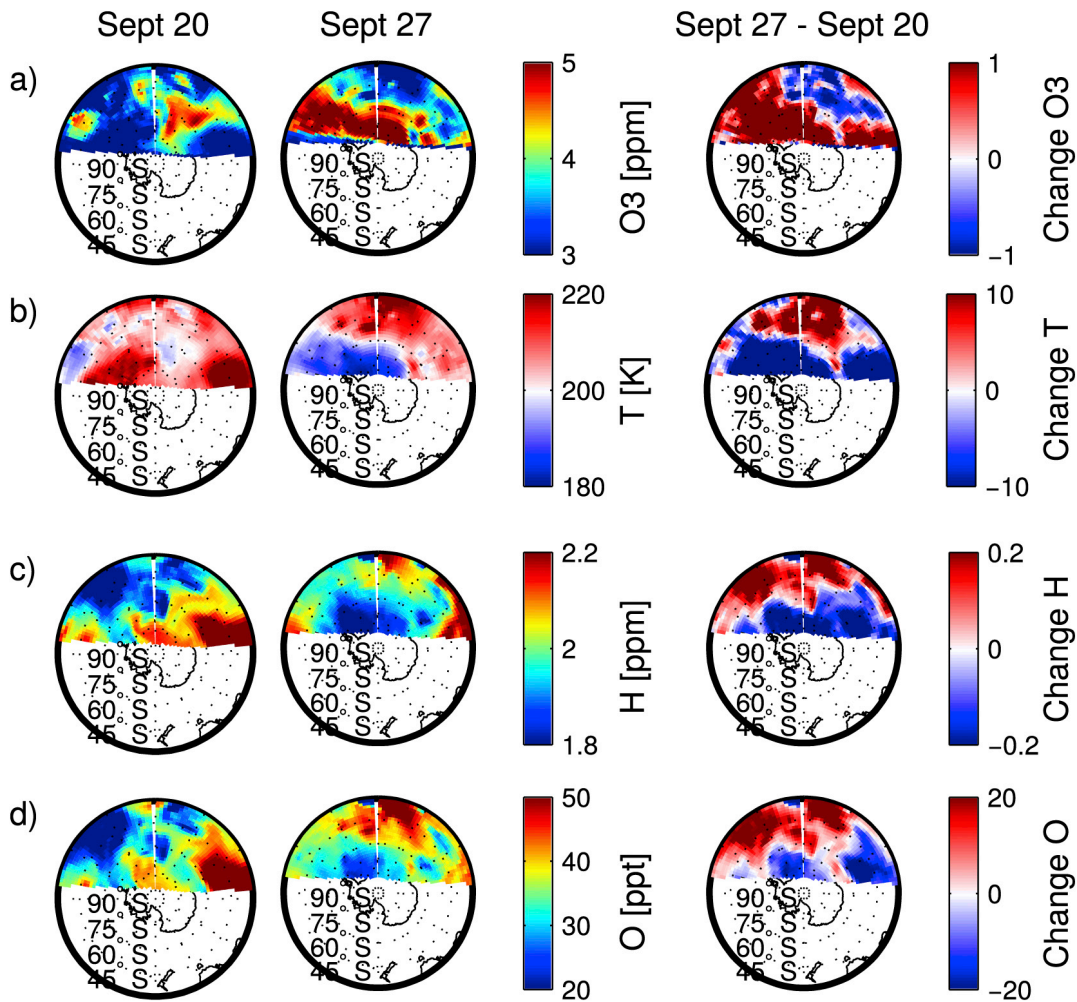

**Fig. 6.** Geographical extent of nighttime: a) ozone (ppm), b) temperature (K), c) atomic hydrogen (ppm), and d) atomic oxygen (ppt) from SD-WACCM. The altitude shown is 85–95 km, where the secondary ozone layer is situated. Left panels show September 20 2002, corresponding to the last peak in eastward wind strength before the SSW (see Fig. 1 for identification of max and min days). The middle panels shows September 27 2002, the day of maximum wind strength in westward direction during the SSW. The left column of panels show the change from September 20 to 27. The outer latitude boundary of the maps is 30°S.

To examine the geographical extent of the ozone increase at the onset of the SSW, we show SD-WACCM polar maps at the altitude of the secondary ozone layer for September 20, a time of a PNJ maximum, and September 27, a time of PNJ minimum (as shown on Fig. 1), as well as their difference (see Fig. 6). Only nighttime is shown, with masked daytime region values. On September 20, higher amounts of ozone are only found in one sector of the night, for latitudes equatorward of 75°S. On September 27, the region with high ozone is covering all longitudes poleward of 60°S. The temperature decrease can be seen also for all longitudes in the polar area, and down to 30°S in some areas. There is overall a good spatial anti-correlation between ozone and temperature. Note that the GOMOS observations, with their limited sampling range, are not optimally sampling the region of high ozone suggested by the model. Both the changes in atomic oxygen and atomic hydrogen follow the temperature change, reflecting the impact of the upwelling lowering their abundances. Other areas with higher amounts of ozone can be seen also further equatorward, down to 30°S at some longitudes, associated with cold temperatures.

Corresponding changes during the three PNJ reductions preceding the major SSW are also shown (see Fig. 7). While the patterns of mesospheric coolings vary somewhat from event to event, the maps highlight the strong spatial correlation between the temperature and atomic oxygen and hydrogen changes, and the spatial anti-correlation between temperature and ozone. Similar responses among these variables during the preceding minor events lend further confidence to the results noted during major warming.

## 4. Conclusions

In 2002 the first ever observed major stratospheric warming occurred in the southern hemisphere. Observations by the Global Ozone Monitoring by Occultation of Stars instrument on the Envisat satellite observe a mean 40% increase of ozone in the secondary ozone layer in the Antarctic polar mesopause region during the month of September. Using the

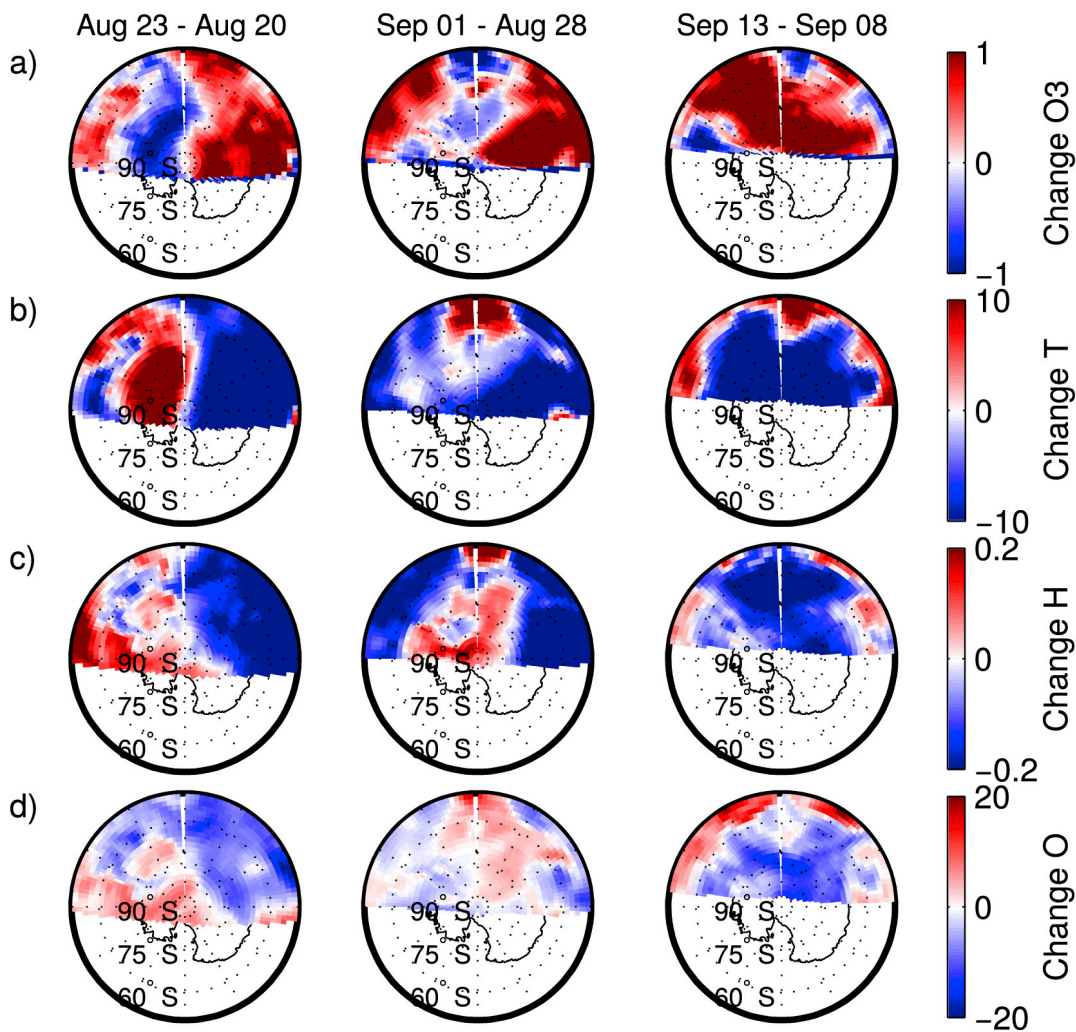

**Fig. 7.** Geographical extent of the change in nighttime: a) ozone (ppm), b) temperature (K), c) atomic hydrogen (ppm), and d) atomic oxygen (ppt) from SD-WACCM, during the three minor warmings. The altitude shown is 85–95 km. The change is from the day of maximum zonal wind to the day of minimum zonal wind strength, as identified in Fig. 1. The outer latitude boundary of the maps is 45°S.

Whole Atmosphere Community Climate Model with specified dynamics, we study the major warming of late September 2002 and several warmings in the preceeding two months. We conclude that enhanced upwelling in the mesosphere leads to a polar mesospheric cooling, and that this cooling is the main contributor to the short-duration peaks in nighttime ozone. The concurrent decrease of atomic oxygen can slightly reduce the effect of temperature change. The study of the nighttime mesospheric ozone during northern hemispheric stratospheric warmings by (Tweedy et al., 2013) attributed the ozone enhancements mostly to the temperature and atomic hydrogen fluctuations. In our case, however, changes induced by the warmings on atomic hydrogen play a minor role, and the seasonal decrease of atomic hydrogen is more influential. This is probably due to the occurrence of the SH events in late winter/early spring, wich is in a different phase of the seasonal cycle compared to the NH mid-winter SSW events.

**Acknowledgments**

CSJ, YOR and FS have been funded by the Norwegian Research Council through project 222390 (Solar-Terrestrial Coupling through High Energy Particle Precipitation in the Atmosphere: a Norwegian contribution). VL is supported in part by the Large-Scale Dynamics Program at the National Science Foundation (NSF) under awards AGS-1642232 and AGS-1624068 and the Kerns Palmetto Professorship supported by the Coastal Carolina University's Provost office. The GOMOS mission was part of the ESA Earth Observation program, and we would like to thank Alain Hauchecorne for providing us with the data. We kindly thank Doug Kinnison at NCAR for the SD-WACCM model output.